# Starlink Mini Satellite Brightness Distributions Across the Sky

Anthony Mallama[1], Richard E. Cole, Jay Respler[1],
Cees Bassa[1], Scott Harrington and Aaron Worley

[1] Centre for the Protection of Dark and Quiet Skies
from Satellite Constellation Interference

2024 January 3

Abstract. The illumination phase functions for Starlink Mini satellites are determined for times of twilight and darkness. Those functions are then evaluated to give apparent magnitudes over a grid of points across the sky and over a range of solar angles below the horizon. Sky maps and a table of satellite magnitude distributions are presented. The largest areas of sky with satellites brighter than magnitudes 6 and 7 both occur during twilight. Brightness surges, known as 'flares', are also characterized.

1. Introduction

Satellites leave streaks on astronomical images which can diminish their scientific usefulness. Some blemishes from spacecraft brighter than magnitude 7 cannot be removed by image processing, so their damage is permanent (Tyson et al., 2022). Satellite constellations are the main concern of professional researchers due to the large numbers of orbiting bodies (Barentine et al., 2023).

Amateur astronomers and others who value the aesthetics and cultural significance of the starry sky are also affected by satellites (Mallama and Young, 2021). Spacecraft brighter than magnitude 6 are visible to the unaided eye as unwanted distractions.

SpaceX operates the largest satellite constellation with more than 5,000 Starlink spacecraft already in orbit (Jonathan McDowell, https://planet4589.org, retrieved 2023 December 21) and regulatory approval for many thousands more. The brightness characteristics for all types of first-generation spacecraft are described in a review article by Mallama et al. (2023). The company began launching their second-generation Starlink spacecraft in 2023. The current type is called Mini because it is smaller than the full-sized Gen 2 satellites which will follow. However, the Minis are four times larger than Gen 1 spacecraft.

This paper reports on the brightness for Mini satellites with an emphasis on their apparent magnitudes. Section 2 summarizes brightness models for two earlier types of Starlink satellites. Section 3 discusses the observational methods used to record data for this study. Section 4 describes the phase function which is used to represent magnitudes. Section 5 gives the results as sky maps and also in a numerical table. Section 6 characterizes short duration brightness surges which are also known as flares. Section 7 lists some limitations of this study and Section 8 presents the conclusions.

2. Earlier brightness models

Cole (2020, 2021) developed a bi-directional reflectance distribution function (BRDF) for the VisorSat type of Starlink spacecraft, a large part of the Gen 1 constellation. The function takes account of the satellite antenna panel and solar array. His model considers eight angles and other factors relative to the spacecraft, the observer and the Sun. Examples are the off-base view angle measured at the spacecraft between nadir and the direction to the observer, and the Sun depression angle taken between the horizontal at the satellite and the direction to the Sun. The model has 10 adjustable parameters including diffuse and specular reflectivity of the antenna panel, and diffuse reflectivity of the solar array. The single output parameter is the modeled apparent magnitude.

The VisorSat model was fit to 131 observed magnitudes of 66 satellites at their on-station heights. Visual data were recorded by Cole, while V-band measurements were obtained from the MMT9 database (Karpov et al. 2015 and Beskin et al. 2017) and those reported in Walker et al. (2021).

His model was evaluated over a rectangular grid representing the sky as shown in Figure 1. The resulting map shows that VisorSats are generally fainter when seen nearer the horizon except for those in the anti-solar direction.

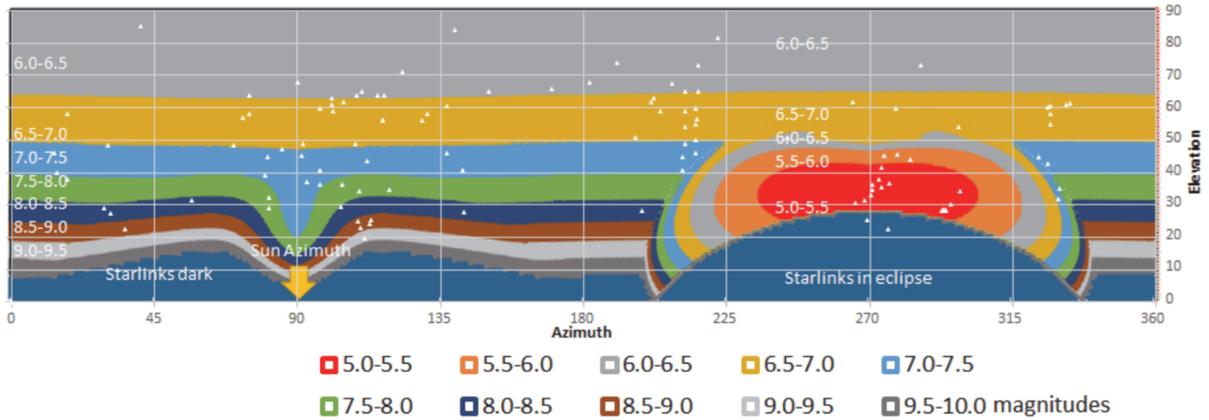

*Fig. 1. Contours of VisorSat apparent magnitudes mapped onto the sky in rectangular coordinates with the Sun 15 ° below azimuth 90 °. Observations are represented by small white symbols. Illustration from Cole (2021).*

Fankhauser et al. (2023) analyzed the Starlink laser communication satellites that followed VisorSat. Their model only required the antenna panel and the solar array. SpaceX provided BRDF functions for these two components based on laboratory measurements. The authors also constrained the BRDF parameters using observed magnitudes in a separate solution. The BRDF models fit the observations better than a simple diffuse sphere model.

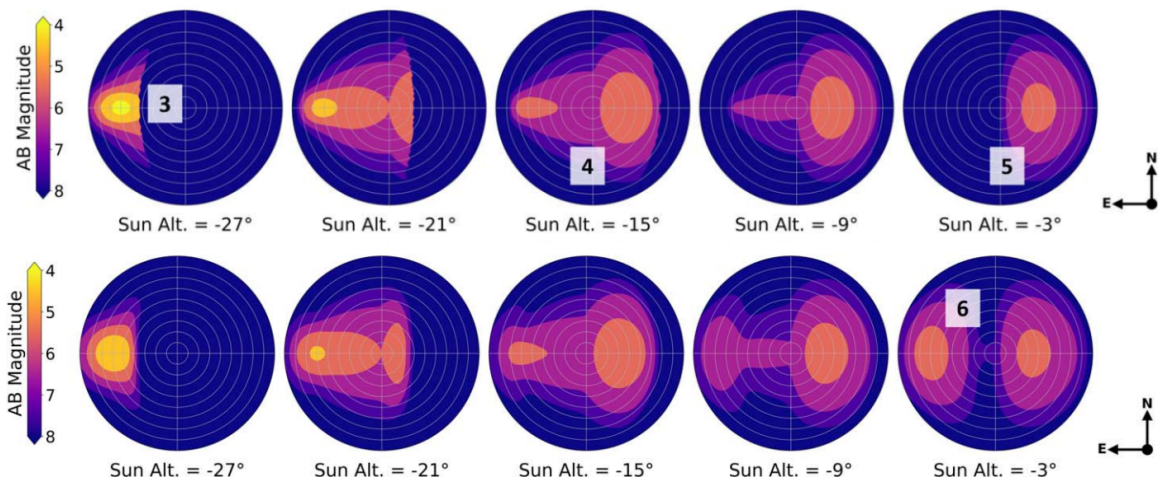

*Fig. 2. The laboratory BRDF model was used to map satellite brightness onto the sky in polar coordinates. The Sun is in the eastward direction at the elevations indicated below each map. Light reflected from the Earth is omitted in the top row and included in the bottom row. Notice the additional brightness in map 6 that does not appear in map 5. This added satellite illumination comes from the Earth's day side. Illustration adapted from Fankhauser et al. (2023).*

Their model was the first to include sunlight reflected from the Earth to the satellite and then down to the observer. This source of illumination adds to spacecraft brightness when they are at low elevations in the solar direction as shown in Figure 2. The authors also point out that this excess luminosity may interfere with searches for potentially hazardous asteroids.

The next two sections discuss the observational methods used to measure brightness for this study and the phase function employed to model magnitudes. Then the results from those observations and modeling are presented in the form of sky brightness maps and a numerical table. The maps are briefly compared to those of Cole (2021) and Fankhauser et al. (2023).

3. Observation sources and methods

Brightness measurements of satellites at their operational heights were acquired by two different means. In the first, observations were downloaded from the database of the MMT9 system (Karpov et al. 2015 and Beskin et al. 2017) on various dates during 2023. This robotic observatory has nine 71 mm diameter f/1.2 lenses and 2160 x 2560 sCMOS sensors. The detectors are sensitive to the visible spectrum and the apparent brightnesses are within 0.1 magnitude of the V-band, based on information in a private communication from S. Karpov as discussed by Mallama (2021). More than 100,000 magnitudes and date/time values were collected from their on-line database (http://mmt9.ru/satellites/) and averaged to give 2,179 five-second means. We computed other quantities needed for this analysis.

Another 418 magnitudes were acquired by visual observers. Many of these were taken at low elevations, where there are few MMT9 observations, in order to sample large and small phase angles. Observers compared the brightness of satellites to nearby reference stars using binoculars and telescopes. Angular proximity between the spacecraft and those stellar objects accounts for variations in sky transparency and sky brightness. Details of the visual method are described by Mallama (2022) and geo-locations for the observers are listed in Appendix A.

4. Phase function

The sky brightness maps shown in Section 2 were based on the BRDF functions described there. This study relies on the phase function which is an empirical characterization of brightness without any reference to the spacecrafts' actual shapes.

Phase angle is the arc measured at the satellite between directions to the Sun and to the observer. This quantity is used to analyze satellite brightness and it leads to the phase function where brightness is the dependent variable of the angle. Brightness in the phase function is the magnitude after it is adjusted to a standard distance of 1,000 km.

Figure 3 shows phase functions for Starlink Mini satellites during nautical twilight, astronomical twilight and darkness. Astronomical twilight corresponds to solar elevations between -12º and -18º. Satellites are brighter when the Sun is farther below the horizon than during twilight because sunlight illuminates the nadir side of the satellite chassis more directly. However, a larger fraction of satellites are eclipsed during darkness than in twilight.

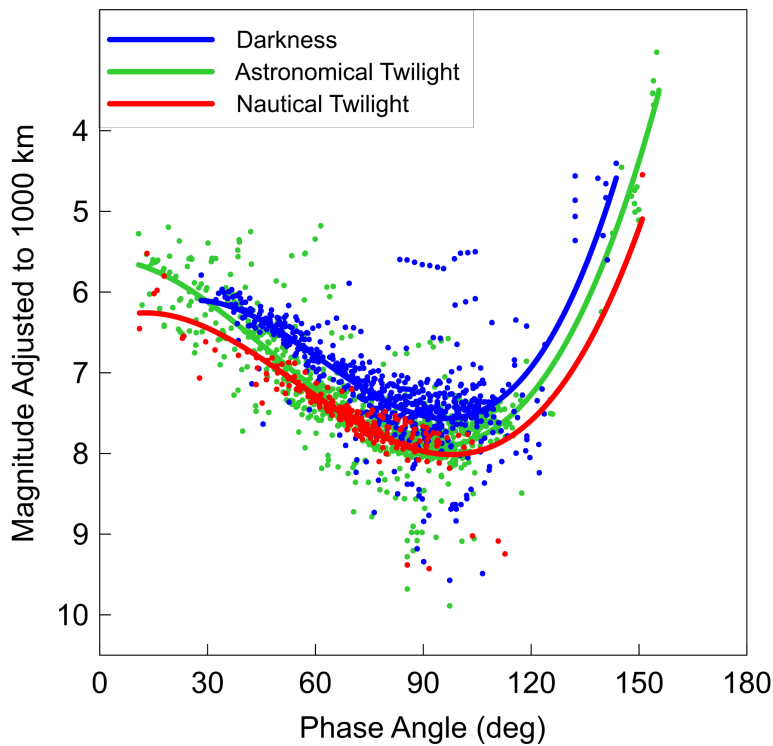

Fig. 3: Individual magnitudes and best-fit phase functions for twilight and darkness.

All three phase functions are characterized by minimum brightnesses near 90º. At smaller phase angles the brightness rises and then levels off. At larger angles the brightness increases steeply. The high luminosity at small angles is expected because the observer is seeing the almost fully illuminated body of the satellite. However, the brightness at large angles must come from forward scattering off the spacecraft base because the Sun is behind it. The coefficients of the cubic fits are listed in Table 1.

Table 1. Coefficients of phase function polynomials

```
     Degree            0           1           2           3
    Twilight*       5.647      -2.672E-5   0.0007799   -5.594E-6
    Darkness        6.657      -0.05474    0.001438    -8.061E-6
   * Nautical and astronomical twilight are combined
```

5. Sky brightness maps and numerical results

The phase functions were evaluated over grids of elevation and azimuth to determine the apparent magnitudes of satellites over the whole sky. Table 2 lists percentages of the sky where spacecraft are brighter than mags 5, 6 and 7 for solar elevations of -12, -18, -24 and -30º. The fractions of the night where these elevations apply depend on the latitude of the observer. One set of columns is for the whole sky while the other is for that part above 30º elevation. Figures 4 through 7 are corresponding maps of the sky where the solar azimuth is taken to be 90º.

Table 2. Percentages of Sky for Limiting Magnitudes

```
    Sol.El.    --- All Sky ---      --- Above 30° ---
               < 5    < 6    < 7     < 5    < 6    < 7
    -12.       2.3   13.6   49.9     0.0   14.9   75.8
    -18.       0.4   10.6   40.7     0.0   15.6   69.7
    -24.       0.0    0.6   14.2     0.0    0.0   16.2
    -30.       0.0    0.0    1.9     0.0    0.0    0.0
```

5.1 Solar elevation -12º

Figure 4 illustrates satellite brightness at solar elevation -12º which is the boundary between nautical and astronomical twilight. There are two areas of sky where satellites are especially bright. One is centered opposite the Sun's azimuth at 50º elevation where magnitudes brighter than 6 occur. The other is in the solar direction near the horizon where brightness exceeds mag 5 over a small area. These regions resemble those derived by Cole (2021) for VisorSats and Fankhauser et al. (2023) for Starlink laser communication satellites.

Spacecraft over half of the sky are fainter than magnitude 7 as shown in Table 2. The '< 7' column under 'All Sky' lists the area where satellites are brighter than mag 7 as 49.9% for the entire sky. However, for spacecraft at greater elevations that value increases to 75.8% as listed under 'Above 30º'. Satellites higher in the sky are at smaller distances which makes them appear more luminous. The regions brighter than mag 6 are 13.6% and 14.9% for the whole sky and elevations above 30º, respectively. Satellites at elevations below 20º in the anti-solar direction (azimuth 270º) are eclipsed by the Earth's shadow.

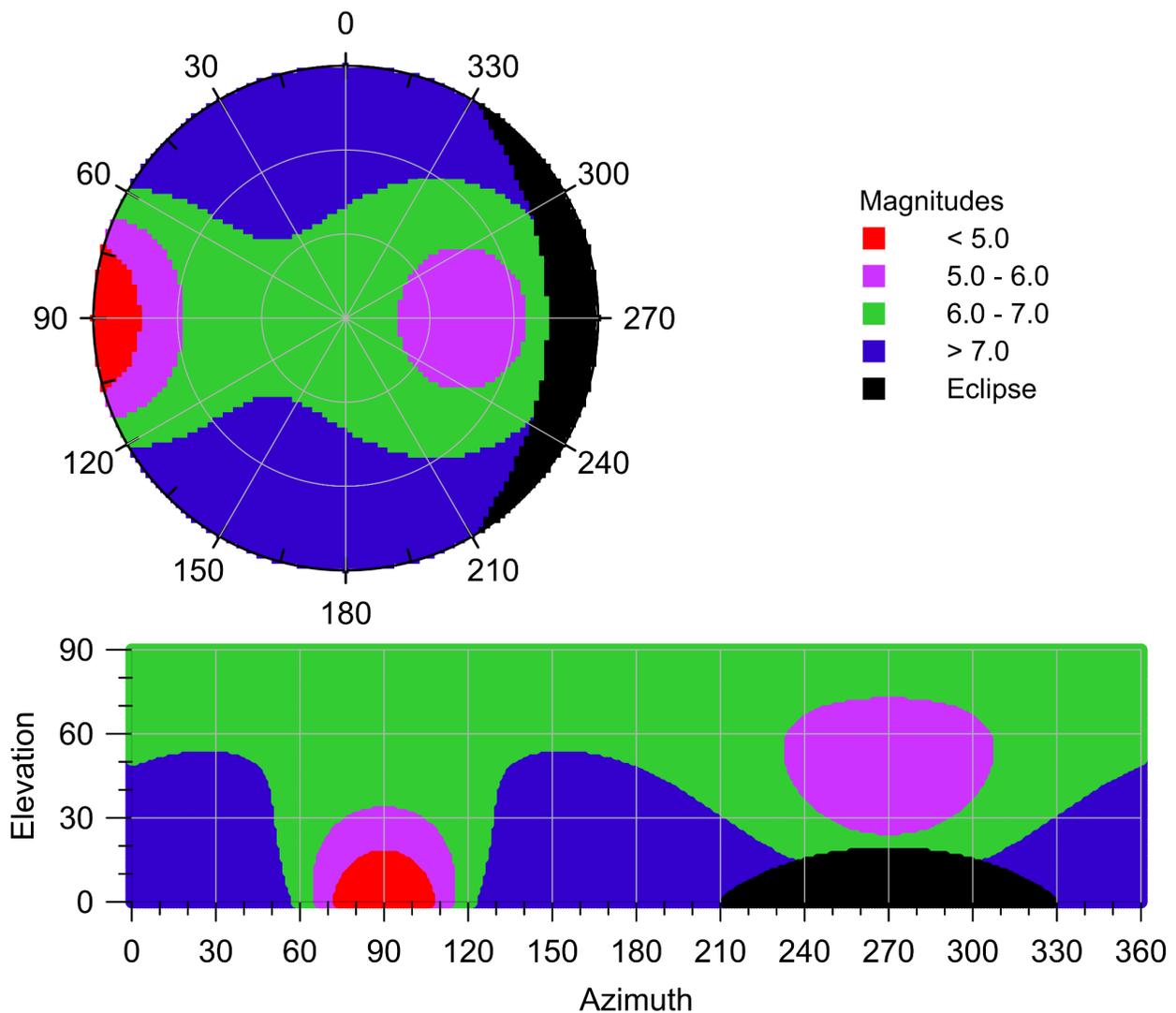

*Fig. 4. Polar and Cartesian sky maps for satellite brightness at solar elevation -12º.*

5.2 Solar elevation -18º

The conditions at solar elevation -18º, which is the boundary between astronomical twilight and darkness, are shown in Figure 5. The eclipse region now extends to 40º at the anti-solar azimuth. The area of magnitudes brighter than 6, centered opposite the Sun at azimuth 270º, approaches zenith.

Satellites brighter than mag 7 cover 40.7% for the entire sky and 69.7% of the sky above 30º. The areas brighter than mag 6 are 10.6% and 15.6% for the whole sky and elevations above 30º, respectively.

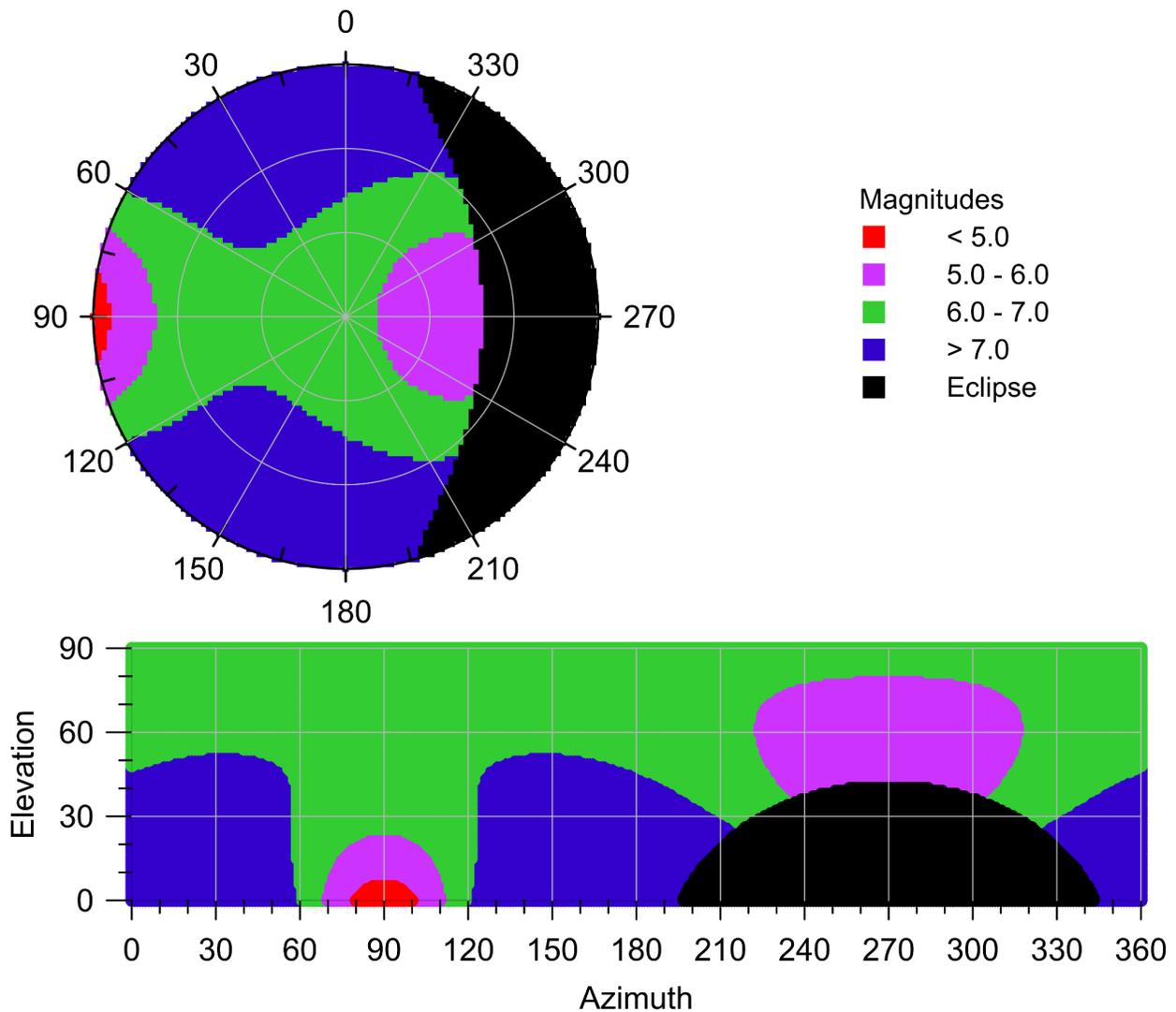

*Fig. 5. Polar and Cartesian sky maps for satellite brightness at solar elevation -18º.*

5.3 Solar elevation -24°

Figure 6 illustrates the conditions during darkness at solar elevation -24°. This occurs about 30 minutes after the end of evening astronomical twilight for low to middle geodetic latitudes. The eclipse region (centered at azimuth 270°) now covers more than half of the sky. Areas brighter than mag 7 comprise only 14.2% of the whole sky and 16.2% above 30° elevation. Those brighter than mag 6 are 0.6% and 0.0%, respectively. A smaller fraction of the sky is impacted by satellites brighter than mags 6 and 7 during darkness than in twilight due to eclipses.

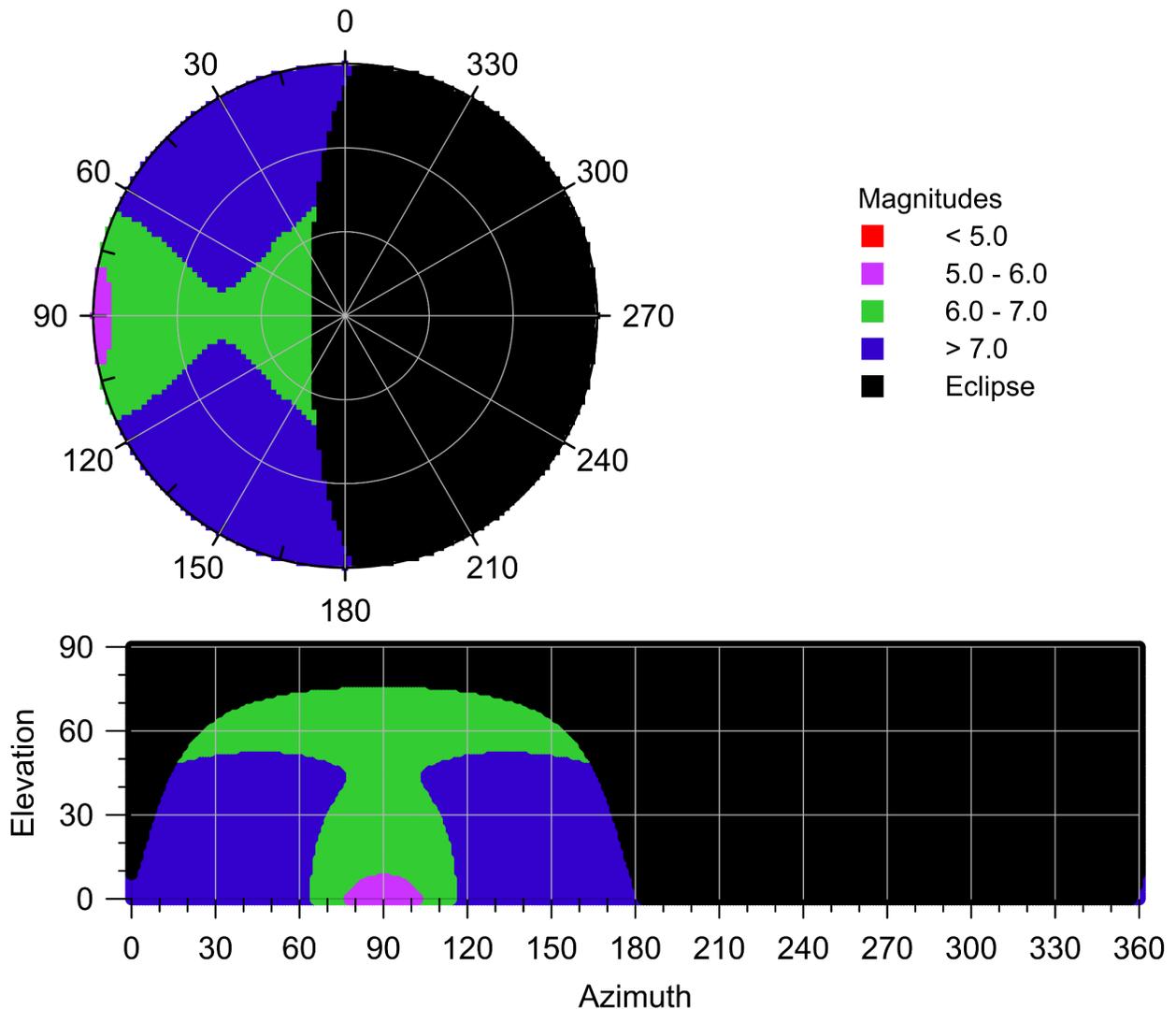

*Fig. 6. Polar and Cartesian sky maps of satellite brightness at solar elevation -24°.*

5.4 Solar elevation -30º

When the Sun is 30º below the horizon, satellites are only illuminated over a small region of the sky extending to about 30º above the horizon in the solar direction (azimuth 90º) as shown in Figure 7. Satellites brighter than mag 7 and 6 cover just 1.9% and 0.0% of the sky area, respectively.

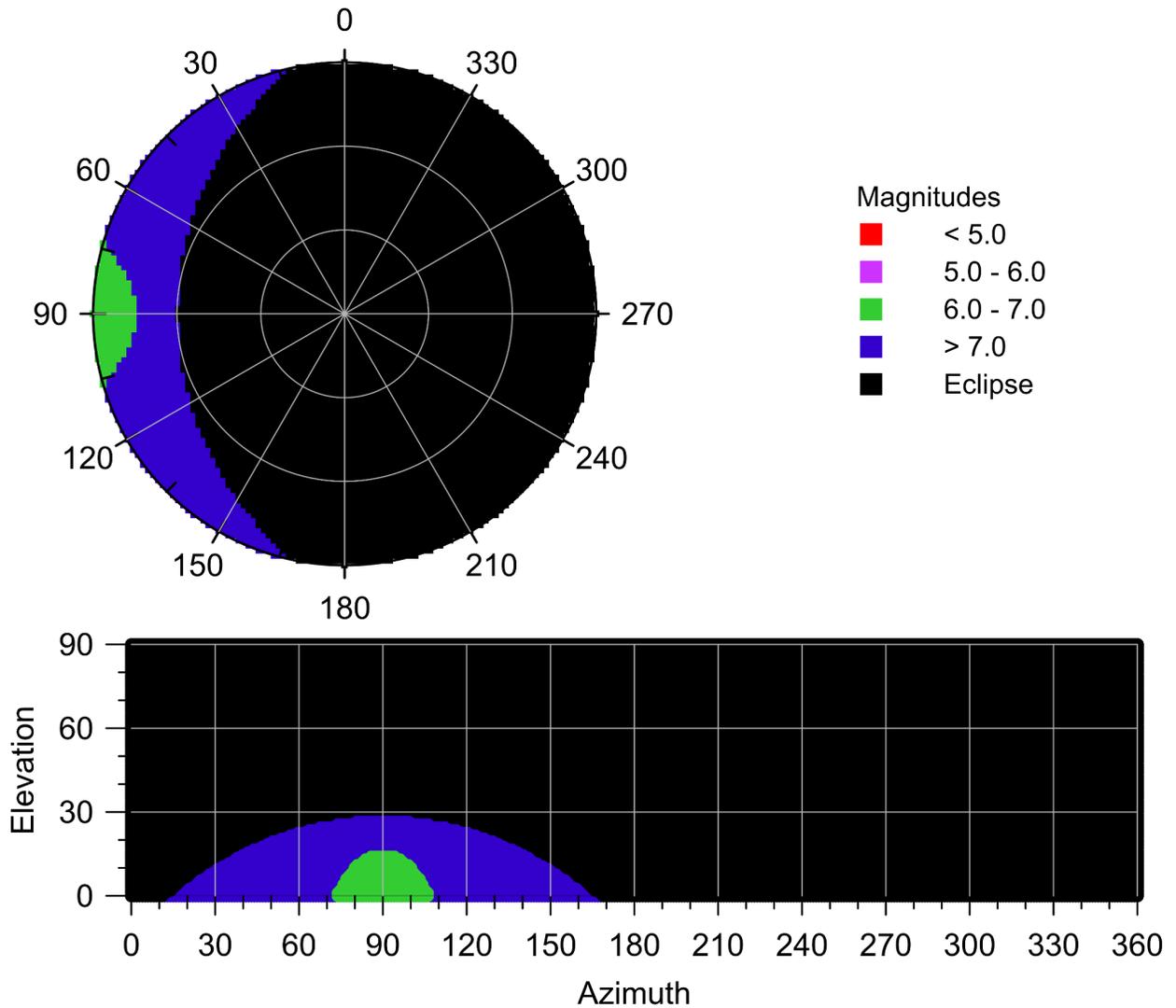

Fig. 7. Polar and Cartesian sky maps for satellite brightness at solar elevation -30º.

6. Flares

The results described above pertain to the ordinary brightness of Starlink Mini satellites which changes slowly during a pass. Such data are well suited to constraining phase function and BRDF models of luminosity. This section addresses short-duration brightness enhancements, called 'flares'. These phenomena are likely caused by pseudo-specular reflections of sunlight from flat satellite surfaces to the observer. An example is shown in Figure 8.

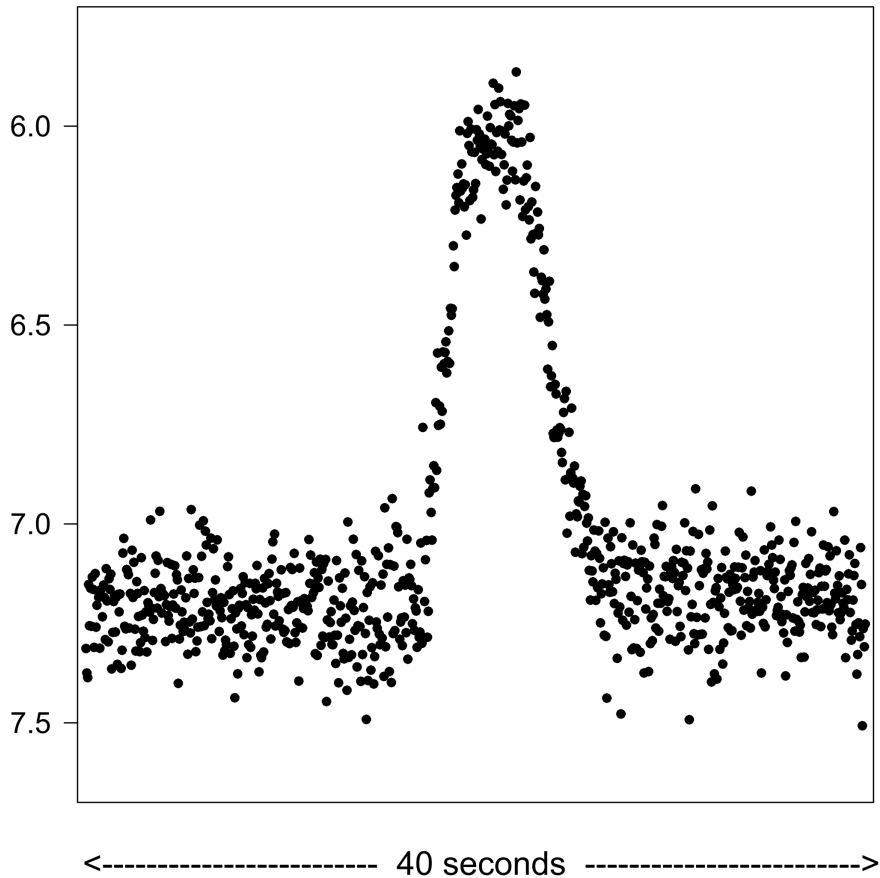

*Fig 8. Starlink-30213 brightened from apparent magnitude 7.2 to 6.0 on 2023 October 25 during a strong flare. The slowly varying baseline is characteristic of non-flaring brightness. These data were recorded at the MMT9 robotic observatory.*

We studied 114 light curves of at least 30 seconds duration each, consisting of approximately 50,000 MMT9 magnitudes in total. Table 3 shows that the Mini satellites exhibited flares with amplitudes greater than 0.5 magnitude during 2.5% of the elapsed observation time and that the mean interval between such flares was 149 seconds. The corresponding values for flares exceeding 1.0 magnitude were 1.0% and 281 seconds. The Table also shows that the flaring characteristics for Mini satellites are about the same as those for VisorSats, while Original design Starlink satellites flared much less often. The results for Visorsat and Original were determined by Mallama (2021) from 100,000 MMT9 observations.

Table 3. Flare characteristics

| Spacecraft Type | Time % by Amplitude > 0.5 | > 1.0 | > 2.0 | Mean Interval (seconds) |
|---|---|---|---|---|
| Mini satellites | 2.5 | 1.0 | 0.1 | 149 |
| VisorSats | 2.8 | 1.0 | 0.1 | 129 |
| Original Design | 0.4 | 0.0 | 0.0 | 622 |

7. Limitations of this study

Precise analysis of satellite luminosity is limited by several factors. One is that brightness variations occur whenever operators modify their spacecrafts' attitudes relative to the Sun and Earth. Such a change was reported for VisorSat spacecraft by Cole (2021) in his BRDF study and by Mallama (2021) as shown in Figure 9.

Mini satellites also undergo brightness variations when SpaceX adjusts their attitudes. This happened when the company initiated a brightness mitigation strategy for spacecraft raising from their initial heights to operational orbits (private communication from SpaceX).

While observations of orbit-raising satellites are not included in this study any attitude changes for on-orbit spacecraft would likewise alter their brightness. The very bright magnitudes near phase angle 90º in Figure 3 are likely due to the satellite being out of brightness mitigation mode.

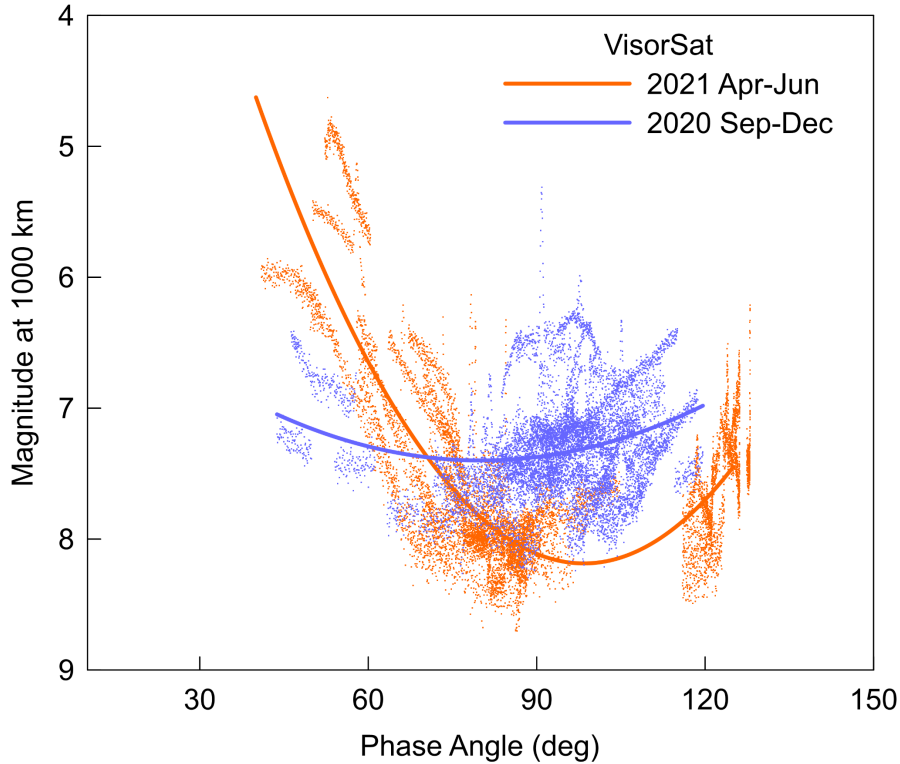

*Fig. 9. VisorSats were brighter in 2021 at phase angles less than 60º and fainter at large angles. Illustration adapted from Mallama (2021).*

Another factor that may affect brightness is the observer's geographic location. Brightness mitigation might be applied more strictly over dark sky locations where the satellites would be most apparent. The spacecraft may be less noticeable in highly populated areas having bright skies.

The two factors described above pertain to variations in satellite brightness itself. There are also limitations in the analysis performed in this study. The phase functions discussed in Section 4 took account of the actual range to the satellites that were observed. However, the sky maps and numerical results reported in Section 5 are based on spacecraft heights taken to be 550 km, although they can be somewhat higher or lower.

The most important limitation of this study may be the use of a phase function to represent brightness. BRDF models as described in Section 2 are potentially a more powerful and precise method. So, comparisons between the results of this study and BRDF models for Mini satellites would be interesting.

Finally, the computed apparent magnitudes are not adjusted for atmospheric extinction. The advantage is that satellite brightness can be compared directly with that of astronomical bodies at the same elevation above the horizon, because they are dimmed by the same amount of extinction. Such results are optimal for evaluating satellites' interference with scientific observations. On the other hand, atmospheric extinction will cause satellites near the horizon to appear fainter to the unaided eye. So, casual observers of the night skies would be less distracted by low elevation satellites.

8. Conclusions

Visual and V-band observations for Starlink Mini satellites were collected and analyzed. Their illumination phase functions were determined for times of twilight and darkness. Those functions were then evaluated over a grid of points across the sky and over a range of solar angles below the horizon.

Maps of satellite magnitude distributions on the sky were presented along with a numerical table of results. The largest areas of sky with satellites brighter than magnitudes 6 and 7 both occur during twilight. The sky maps are briefly compared to earlier models that were developed by other investigators for Starlink VisorSats and the Generation 1 laser communication spacecraft.

Short duration brightness enhancements, called flares, have also been characterized. Flares with amplitudes greater than 0.5 magnitude were present in the light curves 2.5% of the time and the mean interval between them was 149 seconds.

Several limitations of this study are discussed. These include brightness variations attributed to satellite operations as well as those due to analysis techniques.

Acknowledgments

Observations from the [MMT9](#) robotic observatory were highly useful in this study. The [Stellarium](#) astronomy program and the [Heavens-Above](#) satellite website were employed in this research. Two reviewers from the IAU-CPS made suggestions that improved the paper.

Appendix A. Geodetic coordinates of observers

```
Observer            Latitude   Longitude   Ht(m)
R. Cole              50.552     -4.735      100
S. Harrington        36.062    -91.688      185
A. Mallama           38.982    -76.763       43
J. Respler           40.330    -74.445      170
A. Worley            41.474    -81.519      351
MMT9 Observatory     43.650     41.431     2030
```